# Radial profile of the polytropic index of solar wind plasma in the heliosphere

G. Livadiotis
Southwest Research Institute, San Antonio, TX, USA; glivadiotis@swri.edu

Abstract

We combine different measurements of the polytropic index of the proton plasma in the heliosphere: i) near-adiabatic index in the inner heliosphere ~1AU, ii) subadiabatic indices in the outer heliosphere ~20-40AU, and iii) near-zero indices in the inner heliosheath. These observations are unified by a single theoretical model of the polytropic index throughout its radial extent in the heliosphere; the corresponding fitting reveals the decreasing trend of the polytropic index with increasing heliocentric distance $R$. We anticipate that with increasing $R$, (i) the Debye length and mean-free-path decreases; (ii) the Landau damping is less effective, transferring thus less wave energy to particles; and (iii) the collisionality degree increases, indicating that the proton plasma in the inner heliosheath might be collisionless.

1. Introduction

Most space plasmas exhibit positive correlations between density $n$ and temperature $T$, that is, a power-law relationship, $T \propto n^{\gamma-1}$, with $\gamma>1$; the exponent parameter $\gamma$ constitutes the polytropic index, a thermodynamic property characterizing the process of transition of plasmas through thermodynamic states.

Solar wind plasma particles flow throughout the supersonic heliosphere under adiabatic expansive cooling and turbulent heating. Adiabatic thermodynamic processes characterize plasma flows with nearly zero heat transition, and they have $\gamma=\gamma_a$; for three degrees of freedom, $\gamma_a=5/3$. Typically, solar wind near 1AU is quasi-adiabatic, and its polytropic index fluctuates with the most frequent value $\gamma\sim5/3$, i.e., that of the adiabatic process; (e.g., Totten et al. 1995; Newbury et al. 1997; Nicolaou et al. 2014; Livadiotis & Desai 2016). If there were no turbulent heating sources, the radial expansion of solar wind would have been adiabatic and the polytropic index $\gamma\sim5/3$; however, the turbulent heating affects and reduces the polytropic index of the solar wind proton plasma along its expansion throughout the heliosphere.

Turbulent heating increases in the outer heliosphere as the heliocentric distance $R$ increases. Pickup ions (PUIs) contribute to the solar wind heating; this is twofold, as it is caused by the PUI temperature that increases with $R$ (McComas et al. 2017; Livadiotis 2019a), and the PUI turbulent energy (e.g., Smith et al. 2001) also increasing with $R$, but is ~30 times smaller (c.f. Eq.(20) in Livadiotis 2019a).

We combine published measurements and modeling of the polytropic index of the solar wind proton plasma, to understand its radial profile throughout the heliosphere and its consequences in fundamental space plasma processes, e.g., Debye shielding, Landau damping, mean-free-path, and collisionality degree.

## 2. Observations

The polytropic index of the proton plasma in the heliosphere is mostly known for three different measurements: (a) adiabatic index for $R\sim1$AU, (b) subadiabatic indices in the outer heliosphere $\sim$20-40AU, and (c) near-zero indices in the inner heliosheath. In detail:

(a) Many analyses have shown that the polytropic index of the solar wind protons at $R\sim1$AU is on average adiabatic; e.g., Nicolaou et al. 2014, Livadiotis & Desai 2016, Livadiotis et al. 2018, using Wind S/C plasma moments.

(b) The polytropic index of the proton plasma in the outer heliosphere decreases with increasing $R$ between 20-40 AU. Indeed, Elliott et al. (2019) analyzed measurements taken from New Horizons S/C and found that polytropic index decreases with increasing $R$: $\gamma(R)-1\approx-0.0316\cdot(R-30)$; thus, the polytropic index is $\gamma<1$ beyond $R\sim30$-35AU, corresponding to anticorrelation between $n$ and $T$.

(c) Analysis of energetic-neutral-proton measurements, taken from IBEX S/C, showed that the polytropic index is near zero in the inner heliosphere (Livadiotis & McComas 2012; 2013; Livadiotis 2016).

## 3. Results

The independent multi-parametrical model of Fahr & Chashei (2002), which describes the radial profile of the polytropic index of the proton plasma in the heliospehre, is used for unifying and fitting the three different solar wind proton measurements described in §2 (Figure 1),

$$\gamma(R) \cong \frac{5}{3} - \frac{1}{2} \cdot \frac{G_0 \cdot R^{\frac{13}{3}-s} + g_0 \cdot R^{\frac{7}{3}} \exp(-\Lambda_1/R)}{\frac{T_{w0}}{T_S} + \frac{G_0}{\frac{13}{3}-s} \cdot (R^{\frac{13}{3}-s} - 1) + g_0 \cdot \int_{1/R}^{1} t^{-\frac{10}{3}} \exp(-\Lambda_1 t) dt} \quad . \tag{1}$$

The fitting gives the following optimal values: normalized turbulent Alfvén fluctuations $G_0\sim2.2\times10^{-4}$, speed ratio constant $g_0\sim3.0\times10^{-3}$, critical ionization distance $\Lambda_1\sim3.7$AU, exponent $s\sim1$ (corresponding to spectral index $\sim1.4$); solar wind proton temperature ($R=1$AU) $T_{w0}\sim3\times10^{5}$K, Alfven energy expressed as temperature $T_S\sim5.1\times10^{4}$K.

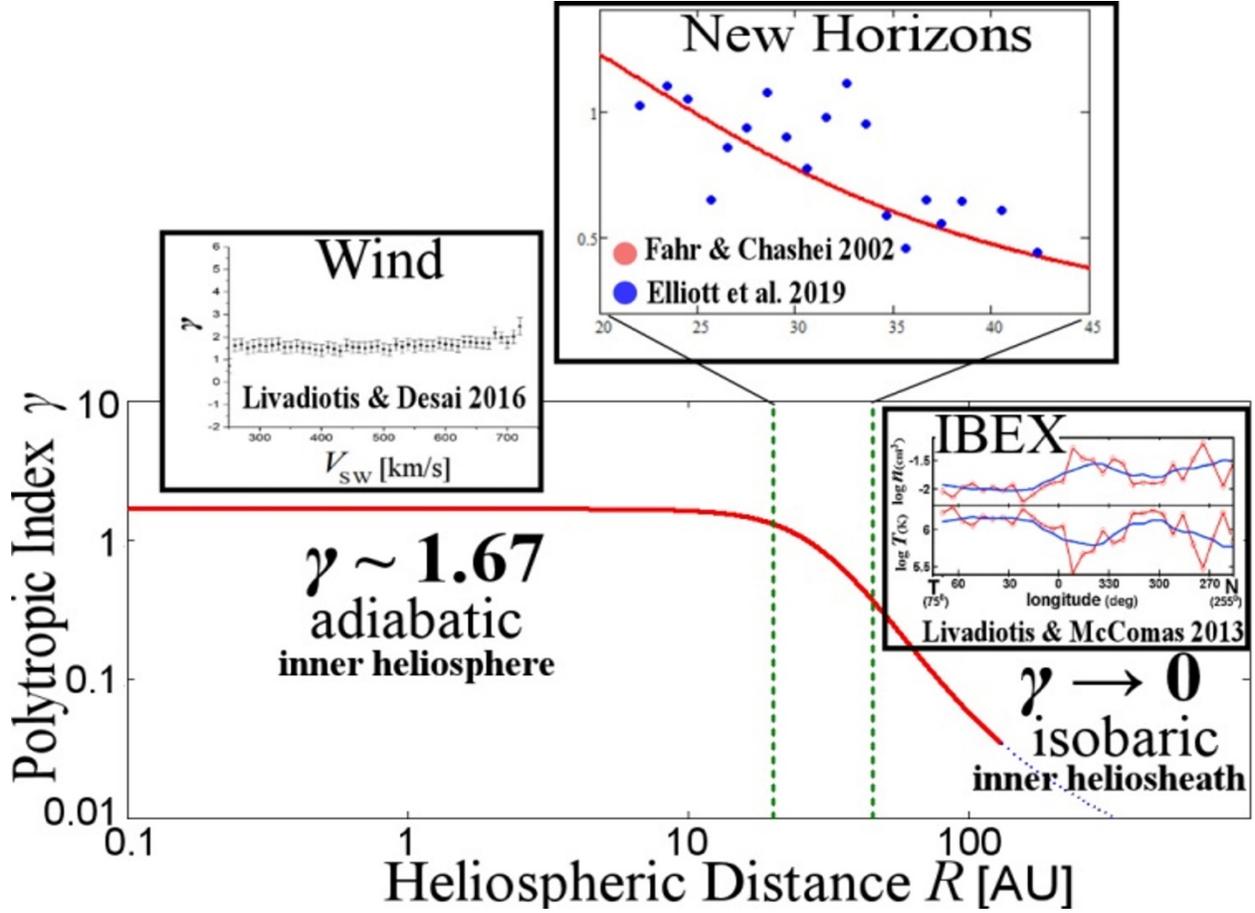

**Figure 1.** The polytropic indices in the (a) inner heliosphere near 1AU (Livadiotis & Desai 2016), (b) outer heliosphere ~20-40AU (Elliott et al. 2019), (c) inner heliosheath (Livadiotis & McComas 2013), are fitted by the model in Eq.(1) (red) (Fahr & Chashei 2002). We observe a consistent connection between the estimated and modeled polytropic indices, leading to a smooth radial profile of the polytropic index in the whole heliosphere, which decreases with increasing $R$.

## 4. Discussion

The continuously decreasing polytropic index with increasing $R$ is critical for space plasma processes.

(1) For the nonzero heating rate $\dot{E}_t > 0$, the polytropic index is smaller than the adiabatic one, following the transport equation (Verma et al. 1995; Vasquez et al. 2007; Livadiotis 2019a),

$$\frac{d}{dR} E_t = \frac{\gamma_a - \gamma}{\gamma_a - 1} \cdot \theta_p^2 \cdot \frac{1}{R} \quad \text{or} \quad \dot{E}_t = \frac{\gamma_a - \gamma}{\gamma_a - 1} \cdot u_{\text{adv}} \cdot \theta_p^2 \cdot \frac{1}{R} \, , \qquad (2)$$

where the heating rate per mass $\dot{E}_t$ and the involved parameters vary with respect to $R$; proton thermal speed $\theta_p = \sqrt{2k_B T_p / m_p}$, temperature $T_p$ and mass $m_p$; the advection speed is given by the wind bulk speed.

In the case of a highly variated polytropic index, it can be shown that Eq.(2), is generalized to

$$\frac{d}{dR}E_t = \theta_p^2 \cdot \frac{d}{dR}\left(\frac{\gamma_a - \gamma}{\gamma_a - 1} \cdot \ln R\right) \text{ or } \dot{E}_t = u_{adv} \cdot \theta_p^2 \cdot \frac{d}{dR}\left(\frac{\gamma_a - \gamma}{\gamma_a - 1} \cdot \ln R\right). \qquad (3)$$

The radial profiles of temperature and polytropic index (McComas 2017) can be combined with Eq.(3) to determine the radial profile of turbulent heating.

(2) The Debye length decreases as polytropic index decreases (Livadiotis 2019b; Saberian 2019), while Landau damping is practically effective on scales comparable to the Debye length (i.e., the damping rate decreases exponentially with increasing the ratio of the wavelength over the Debye length; hence, smaller polytropic indices restrict damping to smaller scales, transferring thus less wave energy to particles.

(3) The polytropic index affects the mean-free-path, and thus, the collisionality of the space plasma. The collisionality degree is described by the thermal-Knudsen number $K_T$, defined by the mean-free-path normalized by the thermal conductivity length; (first derived by: Gurevich & lstomin 1979; then, applied by: Krasheninnikov 1988; Bale et al. 2013; Horaites et al. 2015). For $K_T<0.01$, the plasma is collisional and the heat flux is transferred by thermal conductivity; for $K_T>1$, the plasma is collisionless and the heat flux is transferred by thermal energy advected at thermal speed; for the transition region $0.01<K_T<1$, the plasma is quasi-collisional. When the polytropic index decreases, the mean-free-path also decreases and the plasma collisionality degree increases (Livadiotis 2019c). This result challenges the old-known consideration that the plasma in the inner heliosheath is collisionless (e.g., Zank 2016).

## 5. Conclusions

We combined three different measurements of the polytropic index of proton plasma in the heliosphere: i) near-adiabatic index ~1AU, ii) subadiabatic indices in the outer heliosphere ~20-40AU, iii) near-zero indices in the inner heliosheath. These measurements were unified by a single theoretical model of the polytropic index throughout its radial extent in the heliosphere; the corresponding fitting demonstrated the decreasing trend of the polytropic index with increasing heliocentric distance $R$.

We also anticipated that with increasing $R$, (i) Debye length and mean-free-path also decrease; (ii) Landau damping is less effective, transferring lesser wave energy to particles; and (iii) the collisionality degree also increases, indicating that the inner heliosheath plasma might be collisionless.

Additional work is needed to fully understand the exact radial profile of the polytropic index in the heliosphere, its variation caused by the presence of PUIs, and its abrupt/gradual transition of the decreasing polytropic index in the heliosphere with the near-zero index in the inner heliosheath.

Future analyses may address the following science questions:

What is the exact radial profile of the polytropic index in the full radial extent of the heliosphere?

How does the polytropic index vary along the transition of plasma through the termination shock?

What are the implications in the Debye length, mean-free-path, and collisionality degree?